# Dynamic control of speed and trajectories of active droplets in a nematic environment by electric field and focused laser beam

Mojtaba Rajabi [1,2], Hend Baza [1,2], Hao Wang [1], Oleg D. Lavrentovich [1,2,3,*]

[1]Advanced Materials and Liquid Crystal Institute, Kent State University, Kent, OH 44242, USA
[2]Department of Physics, Kent State University, Kent, OH 44242, USA
[3]Materials Science Graduate Program, Kent State University, Kent, OH 44242, USA
[*] Corresponding author. Email: olavrent@kent.edu

One objective of active matter science is to unveil principles by which chaotic microscale dynamics could be transformed into useful work. A nematic liquid crystal environment offers a number of possibilities, one of which is a directional motion of an active droplet filled with an aqueous dispersion of swimming bacteria. In this work, using the responsiveness of the nematic to the electric field and light, we demonstrate how to control the direction and speed of active droplets. The dielectric response of nematic to the electric field causes two effects: (i) reorientation of the overall director, and (ii) changing the symmetry of the director configuration around the droplet. The first effect redirects the propulsion direction while the second one changes the speed. A laser beam pointed to the vicinity of the droplet can trigger the desired director symmetry around the droplet, by switching between dipolar and quadrupolar configurations, thus affecting the motility and polarity of propulsion. The dynamic tuning of the direction and speed of active droplets represents a step forward in the development of controllable microswimmers.

**Introduction**

Active colloids are self-propulsive units capable of transforming stored or ambient free energy into systematic movement [1-4]. In an isotropic environment, active colloids of both living and inanimate types move along random directions unless their trajectories are biased by gradients of chemicals, temperature, or other cues [5-7]. Liquid crystals, used as a medium for active colloids, offer a much higher control level over the microscale dynamics thanks to their long-range orientational order [4, 8, 9]. In particular, by designing patterns of the nematic director $\hat{\mathbf{n}}$ ($\hat{\mathbf{n}} \equiv -\hat{\mathbf{n}}, \hat{\mathbf{n}}^2 = 1$) that specifies the preferred direction of molecular orientation [10], one can command the polarity and geometry of propulsion trajectories [11-18], mediate transitions from individual to collective modes of propulsion [17] and control the spatial distribution of microswimmers [14, 15, 17].

Recent studies show that a nematic liquid crystal not only directs a microscale motion but could also enable it, as demonstrated by nonlinear electrokinetics [8, 11] and by steady directional propulsion of active droplets dispersed in a thermotropic nematic [18]. In the latter case, a spherical water droplet containing randomly swimming bacteria shows directional motility along the overall director [18]. The motility results from rectification of the chaotic flows inside the droplet by the orientationally ordered exterior. It relies on the symmetry of director distortions set by the perpendicular anchoring of the director at the surface of the droplet. A director field of dipolar symmetry, with a point defect-hedgehog on one side, makes the droplets motile, while a quadrupolar director configuration with an equatorial disclination ring does not [18]. The dipolar symmetry is the ground state of the system in the absence of confinement and external fields [19] but the quadrupolar symmetry becomes prevalent when the sample is shallow [20] or when a strong electric or magnetic field is applied [21-24].

In this work, using the responsiveness of a thermotropic nematic to the external electromagnetic fields, we demonstrate full control over the direction and speed of active droplets by two methods: (i) applying an alternating current (ac) electric field, and (ii) pointing a laser beam at the nematic near the droplet. The electric field applied in the plane of the cell using patterned electrodes changes the propulsion direction by realigning the nematic director and controls the speed by transforming the dipolar director structure into the quadrupolar one. An out-of-plane electric field applied across the cell realigns the director perpendicularly to the substrates and thus reduces the in-plane asymmetry and the droplet's speed. The laser beam locally melts the nematic and switches between dipolar and quadrupolar configurations. The laser beam can also reverse the propulsion direction by creating a hedgehog defect on the side that is intended to lead.

**Materials and methods**

*Active droplets.* We use rod-shaped swimming bacteria *Bacillus subtilis* (strain 1085) of a body length 5–7 μm and a diameter ~ 0.7 μm. The bacteria are initially grown on Lysogeny broth (Miller composition from Teknova, Inc.) agar plates at 35 °C for 12-24 hrs; then a colony is transferred to a Terrific Broth (TB) (Sigma Aldrich) liquid medium and grown in a shaking incubator at a temperature 35 °C for 7-9 hrs. The concentration of bacteria during the growth stage is monitored by measuring the optical density. At the end of the exponential growth, the bacterial concentration is about $c_0 = 0.8 \times 10^{15}$ cell/m$^3$. At this stage, the bacteria are extracted from the liquid medium by centrifugation and added to a nematic lyotropic chromonic liquid crystal (LCLC) to achieve a concentration of 20 $c_0$. The LCLC is a 13 wt% dispersion of disodium cromoglycate (DSCG) (Alfa Aesar), Figure 1A, in TB solution, doped with 0.5 wt% of egg-yolk lecithin (Sigma Aldrich) to stabilize the droplets and to set a perpendicular surface anchoring at the lyotropic-thermotropic nematic interface. Note that the nematic LCLC content of the active droplets is not critical for their propulsive ability. Droplets with bacteria dispersed in water without DSCG still propel in a thermotropic nematic environment and could be controlled by the electric field similarly to the experiments described below. As established previously [18], the LCLC interior produces higher speeds of propulsion, presumably because of a better match between the viscosities of the lyotropic nematic inside and the thermotropic nematic outside the droplet and thus a better momentum transfer.

*Inactive nematic environment.* A thermotropic nematic pentylcyanobiphenyl (5CB) (Merck) is used as the continuous medium, Figure 1B. The dielectric anisotropy of 5CB is positive, $\Delta\varepsilon = \varepsilon_\parallel - \varepsilon_\perp = 10$ (200 kHz) [25], where $\varepsilon_\parallel$ and $\varepsilon_\perp$ are dielectric permittivities measured parallel and perpendicular to the director, respectively. Since $\Delta\varepsilon > 0$, the director prefers to align parallel to the external electric field. The bacteria-containing LCLC is dispersed in 5CB in a volume proportion of 1:50 and vortexed to achieve an emulsion with active droplets surrounded by 5CB.

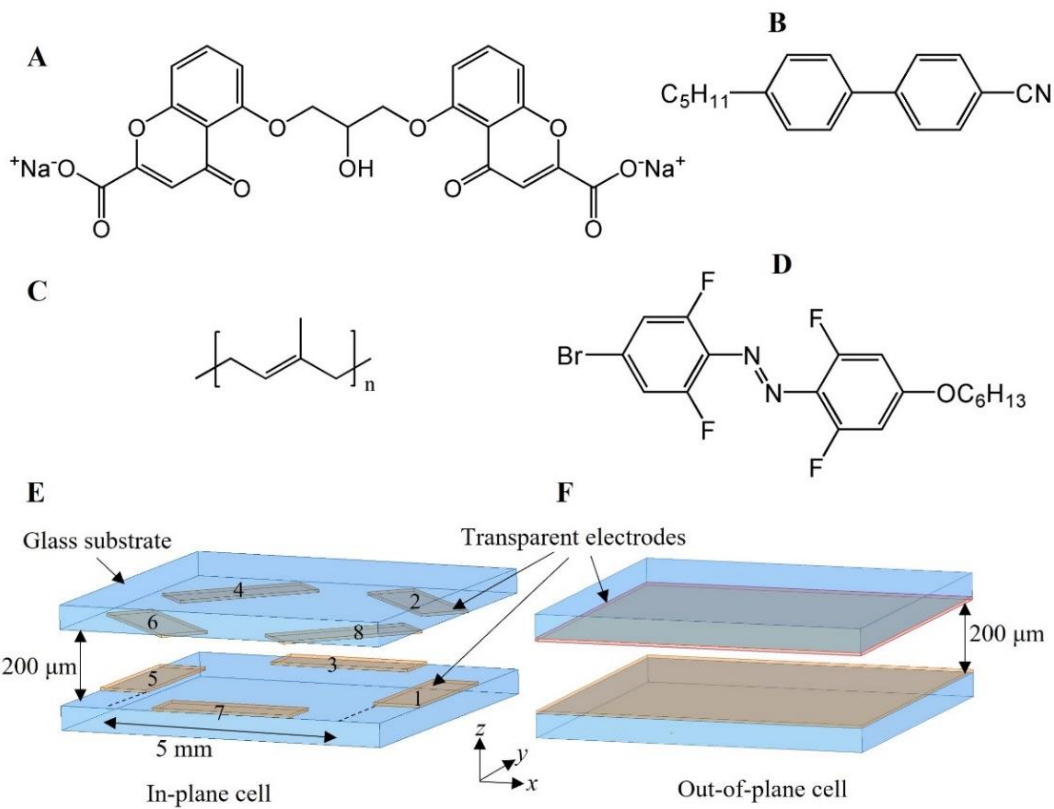

**Figure 1.** Chemical structure of **(A)** DSCG, **(B)** 5CB, **(C)** polyisoprene, **(D)** azodye 4BrOTFAzoO$_6$. **(E)** Geometry of the cell with patterned segmented electrodes. Four electrodes are located on each substrate. The in-plane electric field is applied within the electrodes in the same plate. The scheme is not to scale. **(F)** Geometry of the cell with two transparent electrodes to apply an electric field across the cell.

*Cell preparation.* The emulsion of active droplets in 5CB is filled into cells of a thickness $d = 200$ µm, formed by two parallel glass plates with transparent indium tin oxide (ITO) electrodes at the inner surfaces. The plates are coated with polyisoprene (Sigma-Aldrich) that produces a nearly fully degenerate tangential surface anchoring of 5CB at the glass plates, with a vanishing azimuthal surface anchoring coefficient, $W_a < 3 \times 10^{-10}$ J/m$^2$ [26, 27]. The substrates are prepared by spin coating of the solution of 3 wt% polyisoprene dissolved in methylcyclohexane (Acros Organics) and kept for 30 minutes at 45 °C to evaporate the solvent [28]. All experiments were performed at room temperature within 40 minutes after preparation of the emulsion to make sure the droplets maintain their activity level during the experiment. No sign of degradation or change in the size of droplets was observed within several hours.

*Electrode design and electric field application.* Two different geometries of the electrodes are designed to apply the electric field. (i) A segmented set of eight electrodes is used to apply an in-plane electric field along different directions, Figure 1E. We call this an "in-plane cell". Four electrodes are located at each bounding plate and the separation distance between the opposite electrodes is $l = 5$ mm. The in-plane field is applied between electrodes located on the same plate. To understand the spatial distribution of the electric field, we performed a simulation using Ansys, a commercial finite element analysis modeling software. The simulation shows the electric field applied by two in-plane electrodes is uniform in the central part of the cell, Figure S1 (Supplementary Information). The experiments are performed in this central area (~2 mm$^2$). (ii) The out-of-plane electric field is applied across a cell with a pair of transparent ITO electrodes on the glass plates. We call it "out-of-plane cell," Figure 1F. A Siglent SDG1032X waveform generator and an amplifier Krohn-Hite 7602M are used to apply a sinusoidal alternate current (AC) electric field with a frequency 100-200 kHz to avoid electrohydrodynamic flows. The electric field is increased at a relatively slow rate $\sim 1-3$ V/s, in order to avoid strong backflows. Because of the degenerate azimuthal anchoring, the field realigns the 5CB director in the plane of the cell without hindrance or a memory effect from the substrates. The applied field does not affect the bacterial activity during the experiment time.

*Laser excitation.* An Nd: YVO$_4$ laser (Coherent Verdi-V6) with the wavelength of 532 nm is used to change the director configuration near the active droplets. 5CB is doped with ~1 wt% of azo-dye (E)-1-(4-bromo-2,6-difluorophenyl)-2'-(2',6'-difluoro-4'-(hexyloxy)phenyl)diazene (4BrOTFAzoO$_6$), Figure 1D, which experiences trans-cis isomerization under visible irradiation at 532 nm and thus enhances the sensitivity of 5CB to the laser beam [29]. A prolonged laser irradiation (> 1 min) decreases the bacterial activity, thus in the reported experiments, the irradiation time is kept below 30 s.

The motion of active droplets is observed under an inverted Nikon TE2000 optical microscope equipped with a videocamera Emergent HS-20000C; the trajectories are tracked using the ImageJ software [30].

**Results and discussion**

The surfactant lecithin imposes perpendicular alignment of the nematic 5CB director at the surface of active droplets. The 5CB director around the droplets adopts either a dipolar structure with a point defect called a hyperbolic hedgehog (HH) [19], Figure 2A, or an equatorial disclination ring, referred to as a Saturn ring (SR) [31], Figure 2B. The director $\hat{\mathbf{n}}_0 = (1,0,0)$ far away from an active droplet aligns along the x-axis that represents the direction of the capillary filling during the cell preparation. As described previously [18], in the absence of the electric field, the HH droplets propel along $\hat{\mathbf{n}}_0$ with the hedgehog leading the way. The dipolar fore-aft asymmetric director deformation enables self-propulsion by rectifying flows transferred from the active interior of the droplet to the inactive 5CB environment [18]. By using the dielectric response of 5CB to even weak electric fields, we reorient the director and thus change the droplet trajectory. The trajectory realigns at modest voltages, up to 50 V, without noticeable changes in the speed. At higher voltages realignment is faster but a change in speed might be expected at the same time. To decouple two effects, we redirect the droplet at a voltage $U = 70$ V, in which the response time is moderate, and the speed change is small. Then, we change the speed by applying the field parallel to the director where no redirection occurs.

*Redirecting the active droplets by an in-plane electric field.* An in-plane electric field is applied to 5CB through the segmented electrodes illustrated in Figure 1E. The field realigns the overall director from the initial orientation $\hat{\mathbf{n}}_0$ to a different direction $\hat{\mathbf{n}}_e$ along the field. The active droplets redirect their trajectories to follow the newly-established $\hat{\mathbf{n}}_e$, Figure 2C-E. When the field is switched off, the direction $\hat{\mathbf{n}}_e$ for both the overall 5CB director and the trajectory of the droplet remains intact because of the degeneracy of the azimuthal anchoring at the 5CB-polyisoprene interface.

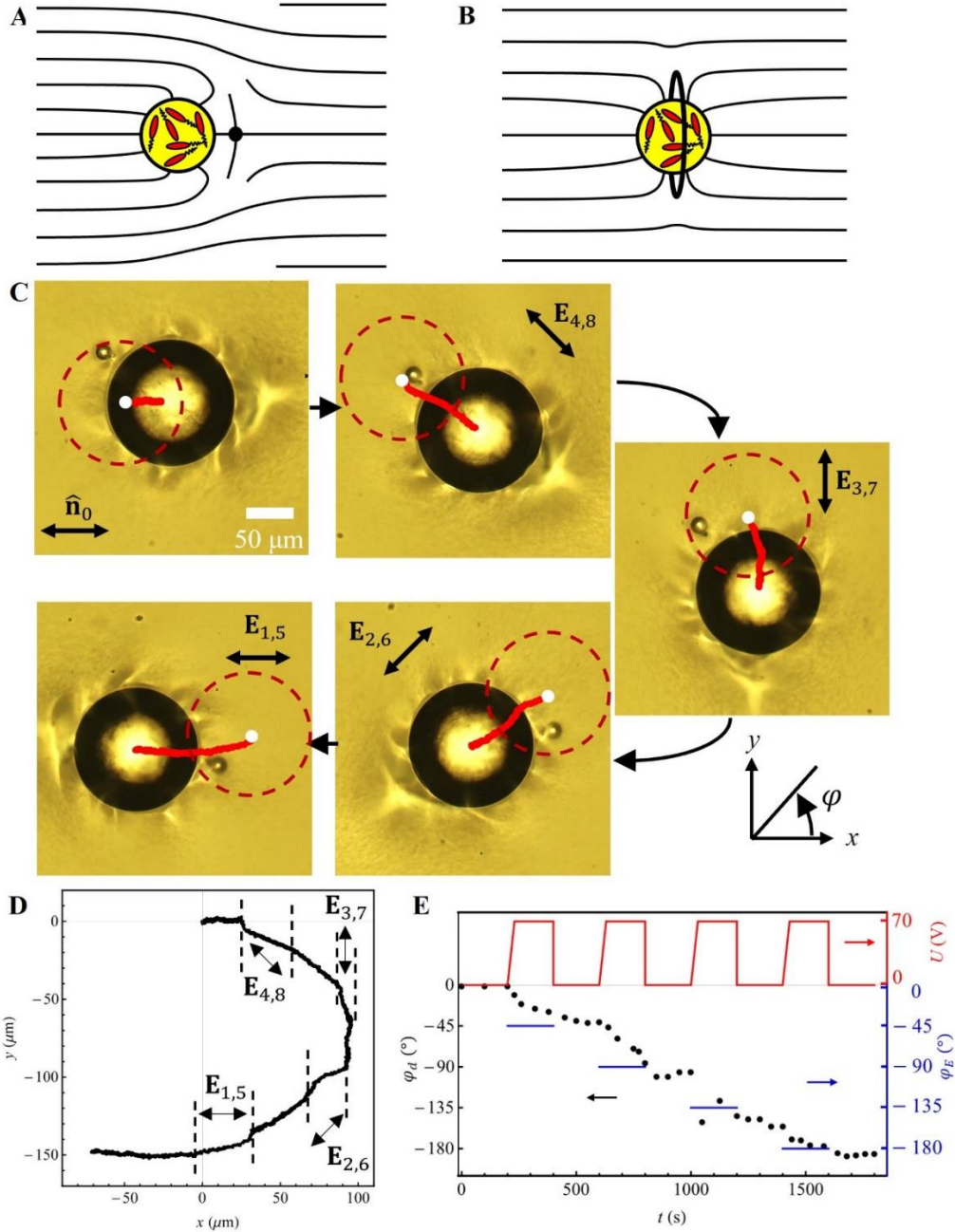

**Figure 2.** Electric field control of propulsion direction of active droplets in the nematic 5CB. Director configuration around a sphere with perpendicular surface anchoring that produces (**A**) a point-defect hyperbolic hedgehog or (**B**) a Saturn ring. (**C**) The sequence of images shows the trajectory of the active droplet. $\mathbf{E}_{i,j}$ means the electric field is applied through the electrodes $i$ and $j$. The white disks show the starting points of the red trajectories, and the red open circles outline the droplet at the starting points. $U = 70$ V, $f = 200$ kHz. (**D**) In-plane trajectory of an active droplet redirected by the field. (**E**) The voltage amplitude (solid line), the angle $\varphi_d$ between the propulsion direction and the $x$-axis (filled circles), and the angle $\varphi_E$ between the field and the $x$-axis (horizontal dashes), all as functions of time.

Because of the sandwich geometry of the cell with two electrodes supplying an in-plane electric field, the actual value of the field is smaller than the simple relationship $E = U/l$ would suggest [32]. Numerical simulation using Ansys software shows that the field in the central part of the cell is reduced by a factor $\beta = 0.8$, i.e., $E = \beta U/l$, so that the applied voltage $U = 70$ V produces $E \approx 11$ kV/m, Figure S1 (Supplementary Information).

The active droplet redirection is defined by the field-induced director realignment. Figure 3 compares the time evolution of the angle $\varphi_d$ that the trajectory of the active droplet makes with the $x$-axis and the angle $\varphi$ that the director makes with the same axis under the same applied field when the droplet is absent. The two dependencies are qualitatively the same and quantitatively similar. We first discuss the dynamics of the director in an applied electric field when the droplet is absent.

Since the azimuthal surface anchoring is negligibly small, the time evolution of $\varphi$ can be obtained from the balance of the electric and viscose torques, $\gamma_1 \frac{\partial \varphi}{\partial t} \approx -\varepsilon_0 \Delta\varepsilon E^2 \varphi$, which assumes director twist by a small angle $\varphi$, but no fluid flow. Here $\gamma_1$ is the rotational viscosity and $\varepsilon_0$ is the vacuum permittivity. Solving the equation with the initial conditions $\hat{\mathbf{n}}_0 = (1,0,0)$ and $\mathbf{E}_{2,6} = E_{2,6}(1,1,0)$, results in

$$\varphi(t) = \frac{\pi}{4}\left(1 - e^{-\frac{t}{\tau}}\right) \tag{1}$$

where $\tau = \frac{\gamma_1}{\varepsilon_0 \Delta\varepsilon E_{2,6}^2}$ is the characteristic director realignment time. The dependency $\varphi(t)$ can be determined experimentally by measuring the intensity of a monochromatic light (548 nm) passing through the cell and two crossed polarizers, polarized along the $x$- and $y$-axes: $I/I_0 = \sin^2(2\varphi)$, where $I_0$ is the maximum value of the intensity, achieved when the reorientation is complete, Figure S2 in the Supplement. The experimentally determined $\varphi(t)$ follows Eq. (1) with a fitted value $\tau = 18$ s, as shown by the dashed line in Figure 3. The fitted value of $\tau$ compares well with the theoretically expected. The material properties of 5CB, $\Delta\varepsilon = 10$ [25], $\gamma_1 \approx 0.14$ Pa$\cdot$s [33], and $E_{2,6} \approx 11$ kV/m ($U = 70$ V), yield $\tau = \frac{\gamma_1}{\varepsilon_0 \Delta\varepsilon E_{2,6}^2} \approx 13$ $s$. The difference between the experimental and theoretical values of $\tau$ could be due to the small-angle approximation in Eq. (1) and due to the difference in the exact temperatures at which the experiments were performed in our study and in characterization of material properties [25, 33].

The active droplet trajectory angle $\varphi_d$ follows closely the evolution of the director angle $\varphi$, Figure 3. The simple model of Eq.(1) thus captures well the essence of the field-controlled droplet redirection by the director realignment. Fitting the experimental dependence with Eq. (1) yields $\tau_d = 22$ s, a value slightly higher than $\tau = 18$ s. The small difference between $\tau_d$ and $\tau$ is expected, since Eq.(1) does not even account for the droplet's presence. The presence of an active droplet leads to the following complications: (i) The swimmers' activity inside the droplet creates flows of the surrounding liquid crystal [18]. (ii) The electric field acting on the liquid crystal medium is different from the value $E = \beta U/l$, because of the high dielectric permittivity of the droplet comprised predominantly of water. (iii) The droplet might experience a higher effective viscosity because of the proximity of the bounding plates [34]. Accounting for the activity-triggered flows, inhomogeneities of the electric field and effective viscosities presents the biggest challenge for further improvements of the theory.

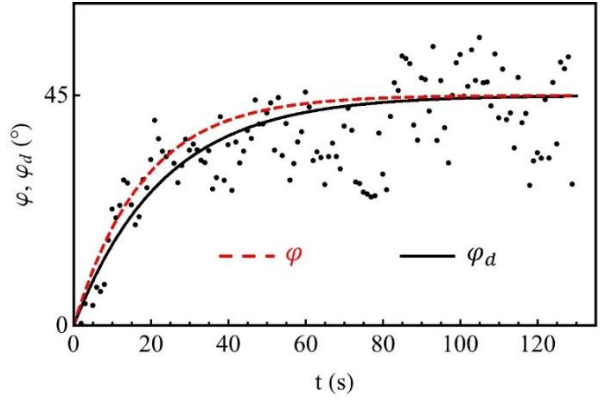

**Figure 3**. The angle $\varphi_d$ between droplet trajectory and the *x*-axis (filled circles), and the angle $\varphi$ between **n̂** and the *x*-axis (dashed line) as functions of time. The solid line is the least-squares fitting of the scattered experimental $\varphi_d$ data. $U = 70$ V, $f = 200$ kHz.

*Speed control by an in-plane electric field.* The in-plane electric field could also control the speed of active droplets by changing the degree of director asymmetry, mainly through the transformation of HH into SR, Figure 4A-C. We quantify the asymmetry degree by the ratio $r_d/R$, where $r_d$ is the distance from the center of the droplet to the plane of the disclination ring. We apply the field parallel to the propulsion direction where it causes no redirection. Increasing the voltage from 0 to $U \sim 50$ V yields a somewhat higher speed, Figure 4B,C ($t = 220$ s), apparently because partial director realignment along the field reduces the effective drag thanks to the viscous anisotropy: the nematic viscosity is smaller for motion parallel to the director. Stronger voltages, 50 V < $U$ < 130 V, reduce the speed, Figure 4B,C ($t = 400 - 800$ s), possibly because the droplet-triggered flows in the surrounding nematic cannot overcome the director aligning action of the applied electric field. At high voltages, $U > 130$ V, the dielectric response of the nematic expands the HH into a disclination ring, Figure 4A ($t = 830 - 880$ s) [21]. During the opening of HH, the speed momentarily increases, as the shift of the opening disclination ring towards the equator means that the center of the droplet shifts in the opposite direction, Figure 4C ($t = 830 - 840$ s). Next, the speed decreases as the disclination ring gradually expands and the director asymmetry $r_d/R$ is diminished, Figure 4A-C ($t = 840 - 880$ s). Eventually, a quadrupolar structure forms, which is incapable of locomotion because of symmetry, Figure 4B-C ($t = 880 - 1200$ s) [18]. Once the electric field is switched off, the disclination ring shrinks to one side of the droplet and the droplet propels again, Figure 4A-C ($t = 1200 - 1590$ s). The direction of the ring's shrinkage and thus the propulsion direction is left-right random with respect to the axis defined by the dipolar structure. The newly created HH could be switched back to the SR, Figure 4A-C ($t = 1590 - 2320$ s).

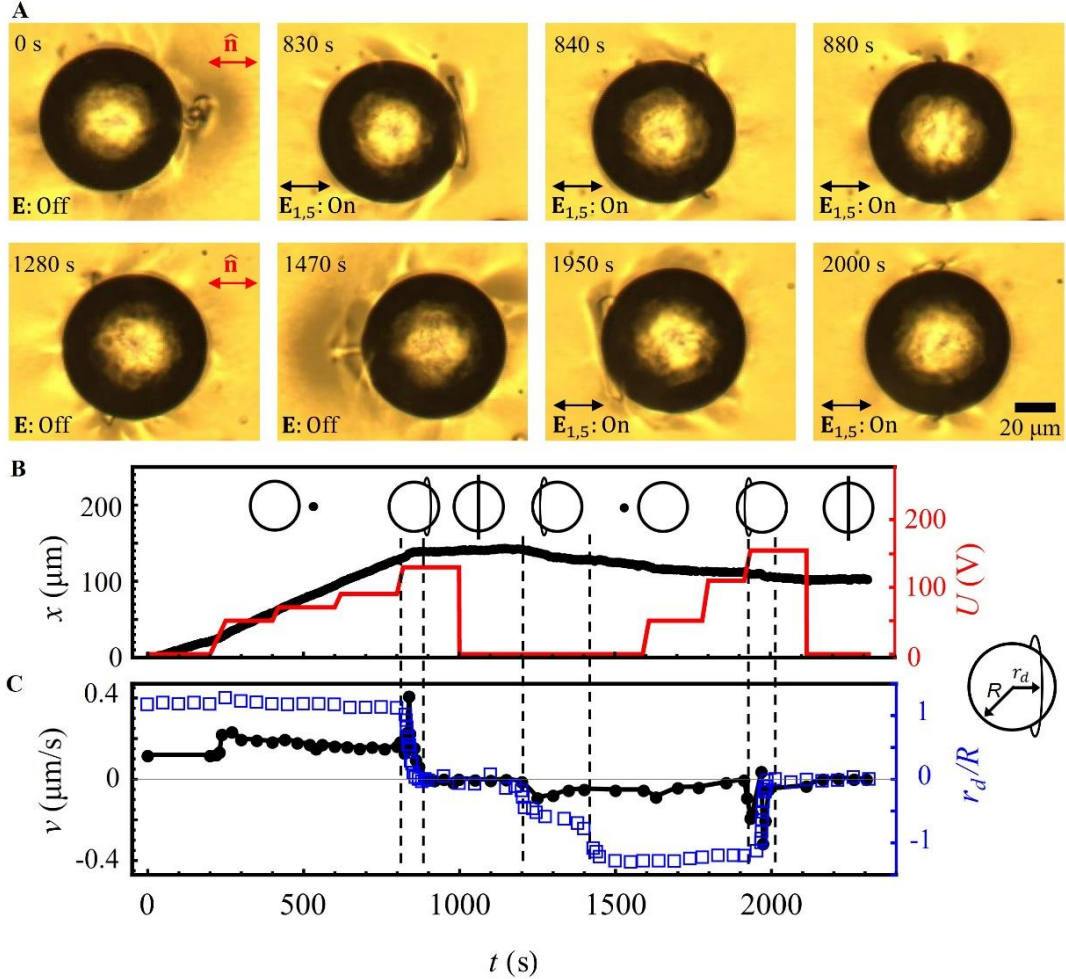

**Figure 4.** **(A)** Transformation of a HH into a disclination ring by an electric field. $U = 130 - 150$ V, $f = 200$ kHz. When the field is switched off the disclination ring shrinks back into the point defect. Switching the field on again results in opening the HH back to SR. **(B)** The horizontal displacement of the active droplet and the applied voltage as functions of time. The inset shows the disclination loop position at different time ranges separated by vertical dashed lines. **(C)** Speed (filled circles) and asymmetry degree $r_d/R$ (open squares) as functions of time. The connecting solid line is a guide to the eye.

*Speed control by an out-of-plane electric field.* The speed of active droplets could also be controlled by the electric field applied across the cell, using transparent electrodes at the bounding plates, Figure 1F. The 5CB director tends to align parallel to the electric field, i.e., normally to the cell's substrates. At small fields, the director far away from the drop remains in the *xy* plane because of the polar anchoring at the bounding substrates, Figure 5A,B. The director realignment around the droplet reduces the asymmetry of HH structure projected onto the *xy* plane and thus decreases the speed, Figure 5A-C. At higher voltages, $U > 4$ V, the far-field nematic reorients parallel to the field, and the droplet loses its in-plane asymmetry and stops, Figure 5A-C. When the field is reduced the overall director realigns towards the *xy* plane, and the droplet resumes its motion.

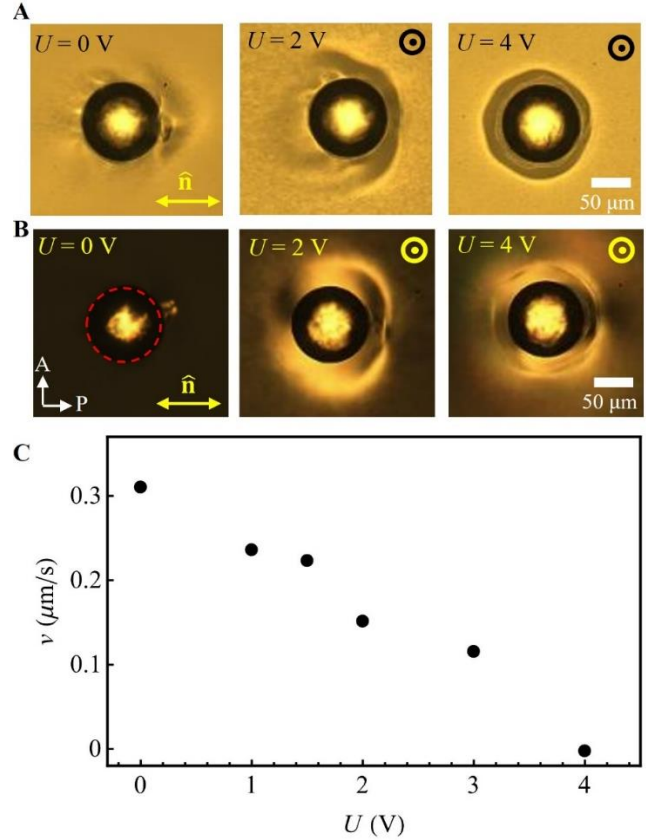

**Figure 5.** Polarizing microscopy observations in **(A)** bright field, and **(B)** with crossed polarizers of active droplets at different applied voltages. The field direction is normal to the cell. $f = 200$ kHz. **(C)** The droplet speed as a function of the applied voltage.

*Controlling speed and polarity of motion by a focused laser beam.* A laser beam can tune the motility and polarity of motion of active droplets. The speed relates to the degree of asymmetry of the director around the droplet. The polarity of motion depends on the location of the HH since the HH leads the droplet. One can establish the desired director configuration, either SR or HH, by locally melting 5CB near the sphere by a laser beam, Figure 6. The laser beam focused on one side of an SR droplet (left side in Figure 6A, $t = 215$ s) causes local melting of the nematic [35, 36], thanks to light absorption enhanced by the dye molecules added to 5CB. When the light is switched off, the isotropic region relaxes back into the nematic state with a web of disclinations, Figure 6A ($t = 230$ s). The disclinations shrink and annihilate with each other until a HH forms on the side irradiated by the laser beam, Figure 6A ($t = 255$ s). Once the HH forms, the droplet starts to propel with the HH leading the way, Figure 6A,B ($t = 255 - 400$ s). One can stop the droplet by transforming the HH back to an SR, by applying the electric field, as described before, Figure 4 and Figure S3 (Supplementary Information), or by focusing a laser beam, Figure 6A. The laser beam is focused at the HH and is moved around the droplet to melt the surrounding nematic, Figure 6A ($t = 400$ s, $415$ s, and $420$ s). When the laser is switched off, the isotropic area transforms to a nematic with multiple disclinations surrounding the droplet, Figure 6A ($t = 425$ s). The disclinations slowly shrink and eventually form a single loop in the form of an SR, Figure 6A ($t = 700$ s). The droplet trajectory fluctuates during the annihilation of the disclinations, Figure 6B ($t = 425 - 700$ s). When the SR of quadrupolar symmetry emerges, the droplet stops, Figure 6B ($t =$ $700 - 830$ s). To reverse the polarity of the motion to the opposite direction, the laser beam is pointed to the opposite side (right side in Figure 6A, $t = 835$ s) of the droplet to form the HH on that side. When the HH is formed the droplet propels while the HH leads the way, Figure 6A,B ($t = 880 - 1000$ s).

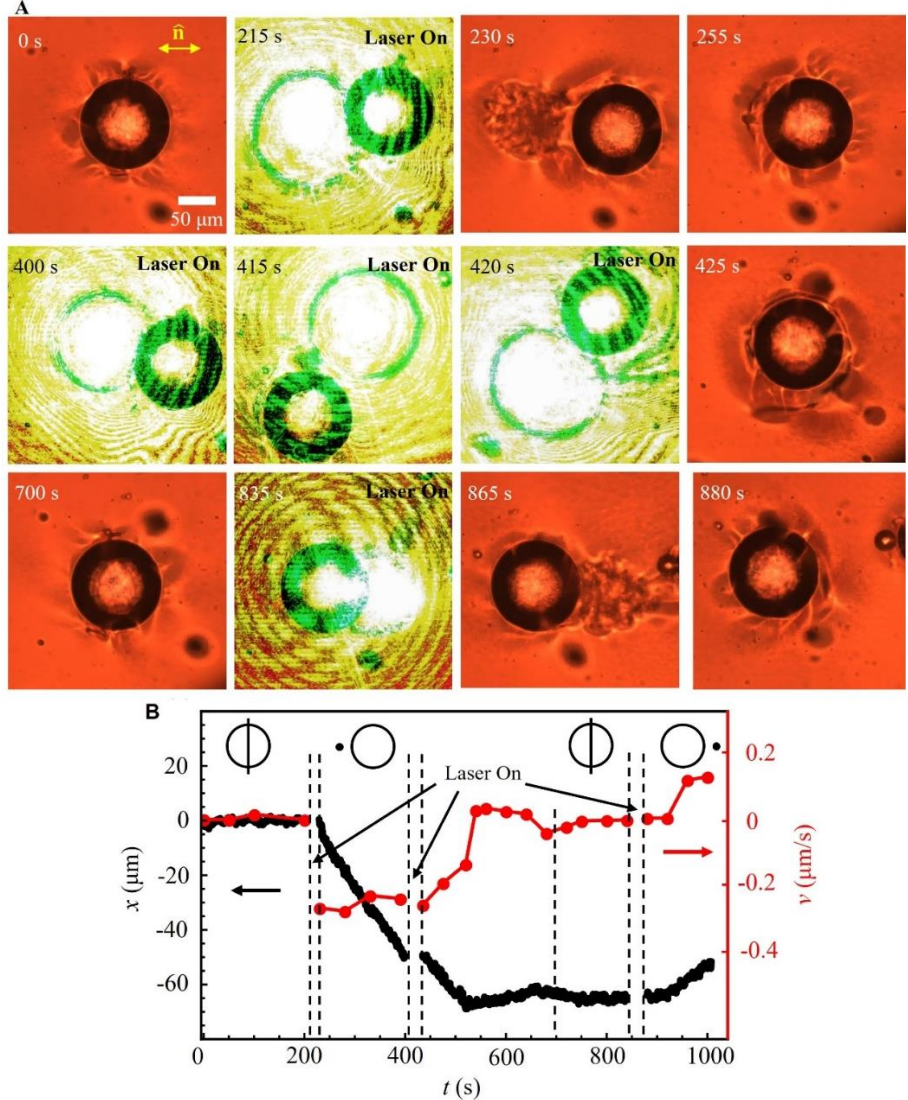

**Figure 6.** Control of speed and polarity of droplet propulsion by light. **(A)** A stationary active droplet with an SR director configuration. A laser beam is pointed to the left side of the droplet and forms an HH on that side. Then, the laser is moved around the droplet at a distance of about $2R - 3R$ from the drop's center to melt the surrounding nematic. When the beam is switched off, the nematic recovers and produces an SR around the droplet. Next, the laser beam is pointed to the right side of the droplet and forms HH on that side and thus reverses the polarity of motion. **(B)** The displacement $x$ and speed $v$ of the active droplet along the overall director of 5CB as functions of time. The droplet with an SR does not propel. After laser beam irradiation and the formation of the HH on the left side, the droplet moves to the left. The laser transforms HH to SR. During the transformation time (425-700 s) the droplet trajectory fluctuates. When the SR forms the droplet shows no propulsion. Finally, the laser is pointed to the right side of the droplet and forms the HH there; the HH leads the droplet to propel to the right.

**Conclusion**

We demonstrated an approach to dynamically control the speed and direction of propulsion of active droplets using an electric field and light. The director field of the nematic environment in which the active droplet is placed realigns parallel to the field and provides a guiding direction for the active droplet. We redirect the droplet by changing the in-plane field direction through a designed set of segmented electrodes. The electric field applied parallel to the overall director controls the speed of the active droplet. In response to the field, the dipolar HH structure transforms into a quadrupolar SR structure with an equatorial disclination ring, which reduces the speed to zero. When the field is switched off, the nematic director around the droplet reconstructs the HH and the droplet resumes a steady unidirectional motion. An out-of-plane electric field also reduces the speed of droplet by realigning the nematic director perpendicularly to the bounding plates and thus reducing the in-plane asymmetry of the HH structure. Using a laser beam, we reversibly transform the immobile SR active droplets into steady propelling HH droplets. We also can change the polarity of motion by first transforming an HH into an SR by a laser or by an electric field and then creating a new HH on the opposite side by a laser beam.

The typical redirection times are on the order of $10^2$ s. These could certainly be shortened, by raising the tunning electric field and using a nematic of a lower viscosity and higher susceptibility to the field.

The observed field-induced HH-SR transformations suggest that the diameter $2R$ of active droplets and the ratio $2R/d$, where $d$ is the cell thickness, are important factors in optimizing the control of trajectories. In the absence of the electric field and in samples with $2R \ll d$, the HH structures are stable when $2R$ exceeds the de Gennes-Kleman anchoring extrapolation length $\frac{K}{W} \sim 10$ μm [18]. Here, $K \sim 10$ pN is the average elastic constant of the liquid crystal and $W \sim 10^{-6} \frac{J}{m^2}$ is the typical strength of polar surface anchoring at the droplet-nematic interface. The droplets with a stable HH configurations show a robust directional propulsion rooted in the dipolar symmetry of the director that rectifies the flows around the droplet [18]. As shown in the present study, these droplets allow one to control their trajectories relatively easy by an electric field and laser beam, since the electric field reorients the director in the entire nematic medium outside the droplet. Small droplets with $2R \sim K/W$ and droplets placed in shallow samples, in which $2R$ closely approaches $d$, are prone to the HH-to-SR transformations [20, 22, 37], which is detrimental for self-locomotion. The quadrupolar symmetry of the SR configurations prevent these droplets from directional propulsion [18]. Furthermore, as found in the previous study [18], droplets smaller than 30 μm, even if they feature an HH director field, do not show rectified flows and thus do not exhibit self-locomotion, apparently because the number of bacteria inside them is not sufficient to produce strong flows. Therefore, the geometrical factors should be accounted for in the design of micromachines employing self-locomotion of active droplets in a liquid crystal.

The advantage of an electric field and laser beams as the means to control the motion of active droplets is their dynamic nature. The previous methods such as patterning of the nematic director were built on a predesigned path approach, where the path of the active colloids was not adjustable once determined [11, 18]. The techniques proposed in our work allow one to adjust the propulsion direction and speed of active droplets in real-time and thus represent a step forward in the design of active and intelligent living matter.

**Conflict of Interest**

The authors declare no conflicts of interest.

**Author Contributions**

MR performed the experiments with the help of HB. HW synthesized the azo-dye. MR performed the numerical simulation. MR and ODL analyzed the data and wrote the paper with the input from all co-authors. ODL supervised the project.

**Funding**

The work is supported by NSF grant DMR-1905053.

**Acknowledgments**

We thank Dr. Sergij V. Shiyanovskii for fruitful discussions and Dr. Jie Xiang for help with the simulations.

# Supplementary Information

## *The spatial distribution of the electric field applied by two in-plane electrodes.*

The field configuration in the cell is simulated using Ansys, a commercial finite element analysis simulator. The parameters are selected to be the same as in the experiment: $l = 5$ mm, $d = 200$ μm, $U = 70$ kV/m, and dielectric permittivity of glass and LC are considered to be 5.5 and 10 respectively. The field is nonuniform near the electrodes but becomes uniform in the central part of the cell, Figure S1A,B.

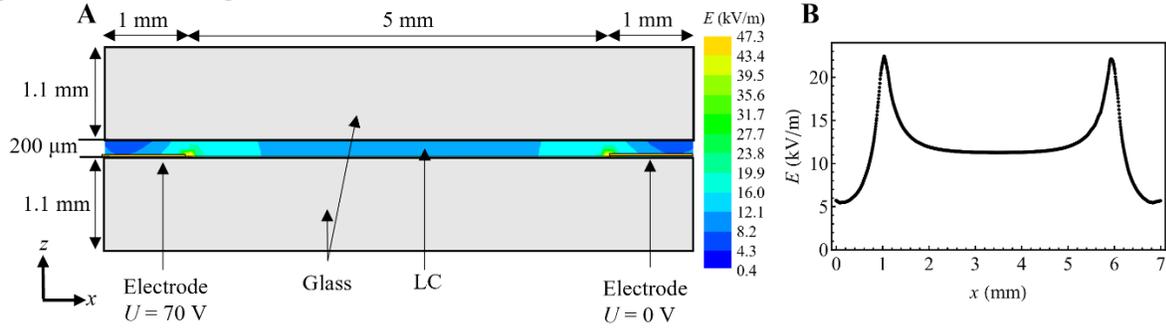

**Figure S1.** **(A)** Simulation of the electric field pattern in a cell with the in-plane electrodes. $U = 70$. **(B)** The amplitude of the field along $x$ at the middle plane of the LC slab.

## *Measuring the director realignment time.*

The director realignment time is measured from the optical transmittance of the LC under crossed polarizers while the director is realigned by an electric field from a dark state parallel to a polarizer to the bright state in which the director is at 45° with respect to the polarizers. The normalized intensity of the transmitted light follows the equation: $I/I_0 = \sin^2(2\varphi)$, where $I_0$ is the intensity when the reorientation is complete. Using Eq. (1) in the main text, the time dependence of the intensity reads $I/I_0 = \sin^2\left(\frac{\pi}{2}(1 - e^{-\frac{t}{\tau}})\right)$. Figure 3A shows the evolution of the normalized intensity $I/I_0$ of the light transmitted through the cell under a continuously applied voltage $U = 70$ V. The experimental data, Figure 3A (solid line) is fitted with the equation, Figure 3A (dashed line), and the fitting parameter is obtained to be $\tau = 18$ s.

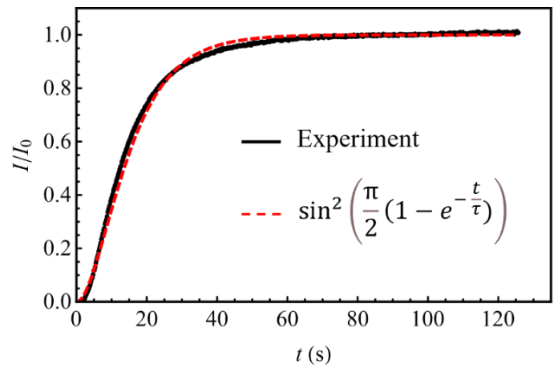

**Figure S2.** Normalized transmitted light intensity of a realigning 5CB under the action of an electric field as a function of time. The solid line represents the experimental data, and the dashed line shows the least-square fitting using $\sin^2\left(\frac{\pi}{2}(1 - e^{-\frac{t}{\tau}})\right)$. $U = 70$ V, $f = 200$ kHz.

*Controlling speed and polarity of motion by a focused laser beam and an electric field.*

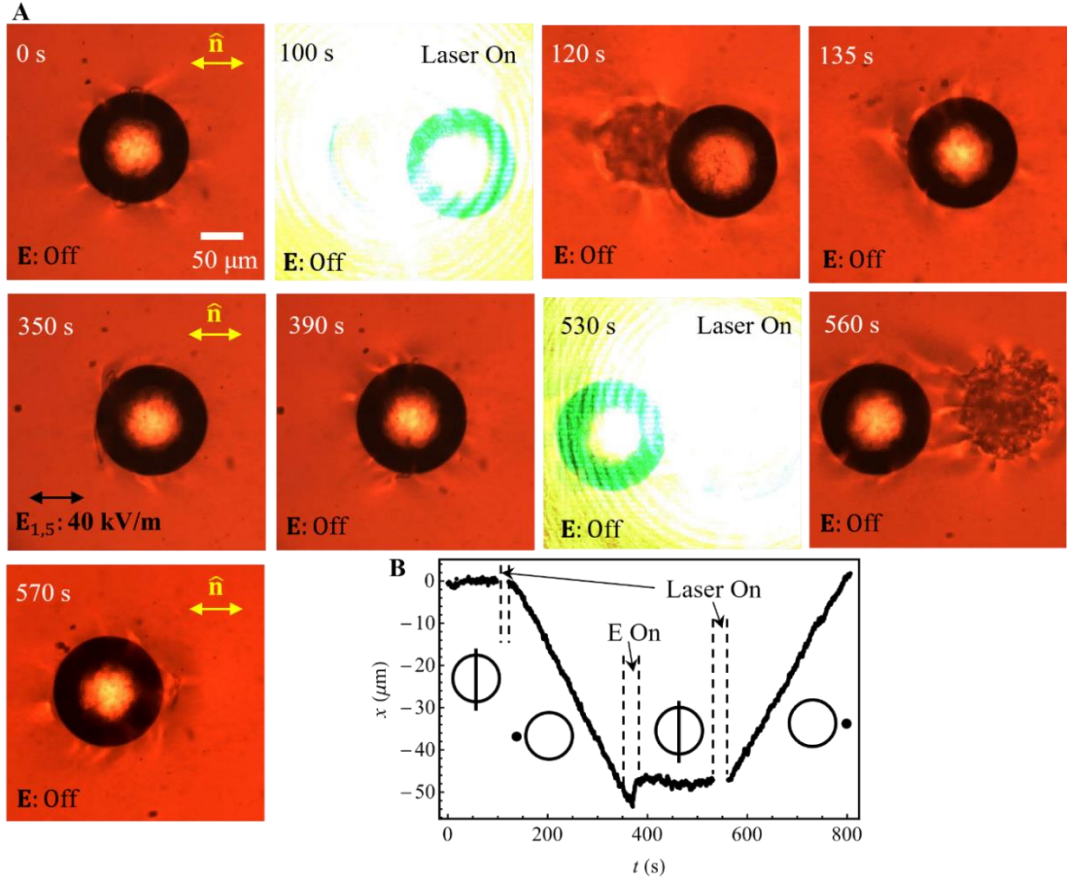

**Figure S3.** Reversing the polarity of droplet propulsion. **(A)** A stationary active droplet with SR director configuration. A laser beam is pointed to the left side of the droplet and forms HH on that side. Thus, the droplet propels to the left. Then, an ac electric field is applied parallel to the $\hat{n}$ and transforms the HH into SR. $f = 100$ kHz. A laser beam is pointed to the right side of the droplet and forms HH on that side and thus reverses the polarity of motion. **(B)** The horizontal displacement of the active droplet as a function of time. The droplet with SR does not propel. After irradiating with laser and formation of the HH in the left side, it moves to the left. Applying an electric field transforms HH to SR again and stops propulsion. Finally, the laser is pointed to the right side of the droplet and forms the HH there, and leads the droplet to propel to the right.

**SI_Video 1**: Directing an active droplet in a nematic environment by changing the direction of an in-plane electric field.

**SI_Video 2:** Reducing the speed of an active droplet by transforming the nematic director configuration from a dipolar into a quadrupolar symmetry using an in-plane electric field applied parallel to the overall director.

**SI_Video 3:** Transforming a quadrupolar director configuration into a dipolar hedgehog structure by a laser beam. The laser melts the nematic near the droplet and when the nematic recovers, it forms the hedgehog in the side that laser is focused. The droplet of quadrupolar symmetry does not propel, while it self-locomotes once the hedgehog forms.